*Article*

# Ranging Performance Analysis in Automotive DToF Lidars

Xiao Guo [1,*]

[1] School of System Design and Intelligent Manufacturing, Southern University of Science and Technology, Shenzhen, Guangdong 518055, China;
* Correspondence: guox2020@mail.sustech.edu.cn;

**Abstract:** In recent years, achieving full autonomy in driving has emerged as a paramount objective for both the industry and academia. Among various perception technologies, Lidar (Light detection and ranging) stands out for its high-precision and high-resolution capabilities based on the principle of light propagation and coupling ranging module and imaging module. Lidar is a sophisticated system that integrates multiple technologies such as optics, mechanics, circuits, and algorithms. Therefore, there are various feasible Lidar schemes to meet the needs of autonomous driving in different scenarios. The ranging performance of Lidar is a key factor that determines the overall performance of autonomous driving systems. As such, it is necessary to conduct a systematic analysis of the ranging performance of different Lidar schemes. In this paper, we present the ranging performance analysis methods corresponding to different optical designs, device selections and measurement mechanisms. By using these methods, we compare the ranging performance of several typical commercial Lidars. Our findings provide a reference framework for designing Lidars with various trade-offs between cost and performance, and offer insights into the advancement towards improving Lidar schemes.

**Keywords:** Lidar ranging performance; avalanche photodiode; silicon photomultiplier tubes

## 1. Introduction

Lidar (Light detection and ranging) is an active remote sensing method that emits laser beams to the target surface and detects the reflected echoes. It subsequently calculates precise distance information by analyzing the flight time of the laser. Lidar was proposed in 1953 [1] and developed rapidly after the invention of laser in 1960s [2]. Laser has been the default light source for Lidar due to its excellent coherence, monochromaticity, directionality and high-power density [3]. Since the introduction of the first Lidar-like system by Hughes Aircraft in 1961 [4], Lidar has been widely used for various applications, and some of them have been commercialized on a large scale [5]. Lidar can detect and range both surface scattering (opaque or immobile) targets [6] and volume scattering (translucent or mobile) targets [7]. In most of the past decades, Lidar was mainly used for military or scientific purposes, such as military reconnaissance [8], aerospace exploration [9,10], air-/ground-/space-borne atmospheric, geographical and oceanographic remote sensing [11]. Recently, low-cost commercial Lidar systems have emerged for 3D mapping and modeling in fields such as intelligent robotic perception [12,13], augmented reality [14], forestry [15], architecture [16], and archaeology [17]. In comparison to these applications, the automotive industry imposes more stringent requirements on Lidar technology [18]. In 2007, DARPA grand challenge demonstrated the great potential of Lidar in sensing surroundings and navigating autonomous vehicles through complicated terrains [19,20]. Autonomous vehicle perception systems usually consist of a combination of active and passive sensors, i.e., millimeter-wave radar, Lidar, ultrasonic radar, and cameras [21,22]. Lidar possesses unparalleled advantages over cameras and millimeter-wave radar in terms of ranging accuracy and resolution, marking a significant milestone in the realm of autonomous driving technology. As depicted in Fig. 1(a), the Lidar system mounted on an autonomous vehicle detects nearby cars by emitting an infrared laser beam and capturing the scattered light reflected from them. The ranging module of the Lidar measures the distance to the target in each direction, while the imaging module scans all directions within its field of view. Consequently, the Lidar generates a 3D point cloud representation of the car, as illustrated in Fig. 1(b). This collection of high-precision and high-resolution point cloud data facilitates more efficient and accurate performance of tasks such as target detection and speed estimation in autonomous driving systems.

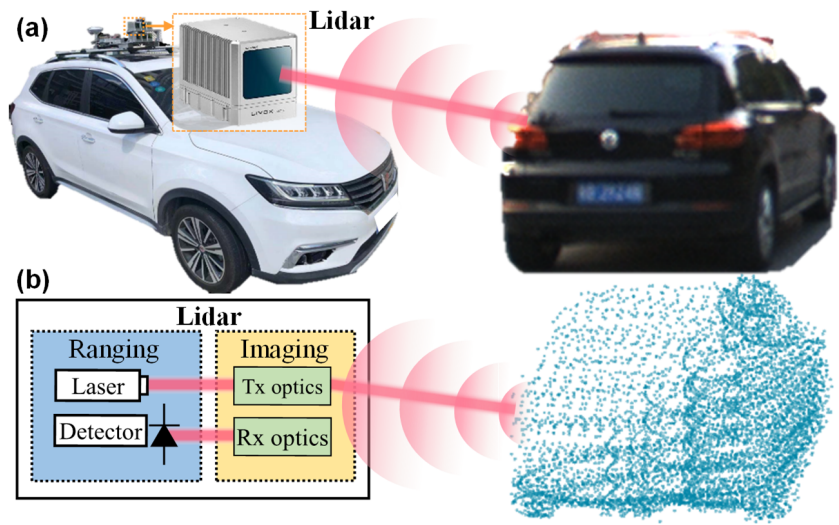

**Figure 1.** Automotive Lidar in an autonomous car. (**a**) Lidar mounted on an autonomous car emits infrared laser light and receives laser scattered by the target (car). (**b**) Basic structure of Lidar. The ranging module of Lidar obtains the distance of the target according to the flight time of laser and the imaging module of the Lidar traverses the direction in field of view. The high-precision and high-resolution point cloud data of the target (car) obtained by Lidar is shown on the right.

In the market, there are numerous schemes for Lidar systems, and the ranging performance, a core metric of Lidar, involves knowledge spanning multiple domains. Literature has explored various aspects including Lidar equations [23-25], detector [26,27], laser technology [28,29], detection methods [30,31], hardware circuit [32,33], architectures [34], among others [35-37]. These studies are confined to single domains, and academia lacks a comprehensive and operationally robust study on the ranging performance of Lidar systems aligned with current industrial standards. This indicates that different manufacturers of lidar systems employ diverse measurement and evaluation methods to assess their nominal ranging performance. This brings considerable confusion to users of lidar systems. Therefore, this paper aims to propose a comprehensive and operationally robust analysis model and method for evaluating the ranging performance of lidar for the industry. Specifically, accurately calculating the maximum detectable range necessitates a dependable estimation of the Lidar's Signal-to-Noise Ratio (SNR). This paper presents a detailed quantitative analysis method for Lidar signal and noise, covering the analysis of Lidar equations, characteristics of laser emitter components and photoelectric detector components, waveform acquisition schemes, and signal processing circuits. The paper is organized as follows: Section 2 provides a detailed introduction to Lidar technology. Section 3 presents a comprehensive Lidar ranging performance analysis model. Section 4 establishes Signal-to-Noise Ratio (SNR) models for two Lidars using different detectors. Section 5 discusses the comparative results of the ranging performance of different Lidars. Finally, Section 6 summarizes the main conclusions drawn from this work.

## 2. Materials and Methods

### 2.1. Lidar technology

Lidar obtains distance based on the time-of-flight between laser emission and return. The basic principle of Lidar ranging comes from the following equation:

$$R = \frac{1}{2} c \cdot \tau \tag{1}$$

where $R$ is the distance of the target, $\tau$ is the time of the round-trip of laser propagation, and $c$ is the light speed. Here we assume the Lidar is working in air and light speed in air approximates light speed in a vacuum.

In Fig. 2(a), the detailed structure of the Lidar is shown. The modulator is responsible for modulating the laser waveform according to various ranging principles. There are three different mechanisms for obtaining time-of-flight based on different laser modulation methods: direct time-of-flight (DToF), amplitude modulated continuous wave (AMCW or indirect TOF) [38], and frequency modulated continuous wave (FMCW) [34]. The measurement principles of these methods are illustrated in Fig. 2(b). The DToF method obtains time-of-flight by directly recording the emission and return times of a short (~ns) pulsed laser. This method is currently the most popular. The AMCW method calculates the time of flight by measuring the phase difference between the received echo and the emitted laser, which is a continuous laser with amplitude modulation in a certain periodic form. While AMCW is easier to implement, it has a limited measurement range, and is therefore often used in blind-filling Lidar. The FMCW method modulates the laser frequency by adjusting the configurations of the laser. Fig. 2(b) shows a popular FMCW scheme with frequency linear modulation. The propagation of laser causes a frequency difference $\Delta f$ between the echoes and the emitted laser. This frequency difference can be obtained by coherent measurement, and has a relationship with time-of-flight as follows:

$$\tau = \frac{\Delta f}{\gamma} \tag{2}$$

where $\gamma$ is the chirp rate of the linear modulation of the laser frequency. In the FMCW Lidar, non-coherent light sources such as background light and potential interference from other devices do not affect its performance. However, the DToF and AMCW methods are susceptible to such influences. The infrared laser source employed by the Lidar can be either an edge-emitting laser (EEL), a vertical-cavity surface-emitting laser (VCSEL), or a fiber-optic laser. EEL laser is the mainstream choice because of its mature supply chain system and extremely high power density. VCSEL has great development potential because of its advantages in narrow line width, low wavelength temperature drift coefficient, good beam shape, easy deployability and low cost. Fiber lasers are usually used in 1550 nm Lidar whose max permissible exposure is higher than that of 905 nm Lidars for the human eye. Lidar can be divided into two categories according to different illumination methods: beam scanning Lidar and flash Lidar. Beam scanning Lidar measures the distance of a point at one time with concentrated laser energy. Generally, scanning Lidar is used as the main sensor of the car for long-distance detection; however, it inevitably causes motion distortion of the point cloud results. Flash Lidar measures the depth of the entire area at one time without motion distortion; nevertheless, it is limited by laser safety regulations [39] and has a low laser power for a single direction resulting in a small measurement range. The receiving part of flash Lidar requires the use of a high-performance array photodetector chip which is still immature. Currently scanning Lidar has been commercialized and many beam scanning methods have been adopted. As shown in Fig. 2(c), Lidar can be classified into mechanical Lidar, hybrid solid-state Lidar, and solid-state Lidar depending on the structure of their imaging module. The evolution of Lidar from mechanical to solid-state Lidar aims for higher reliability, lower cost and smaller size and weight. Mechanical LiDAR refers to LiDAR where the entire ranging module participates in scanning motion. Currently, it is mainly used in single-line LiDAR for robotic vacuum cleaners. Hybrid solid-state LiDAR refers to LiDAR where the ranging module is fixed, and the optical path is scanned through mechanical movement of reflecting or refracting optical components. It is now widely used commercially. However, these methods require mechanical bearings or similar support devices, which may lead to fatigue failure. For most LiDAR systems with mechanical scanning, reliability is undoubtedly a challenge. Solid-state LiDAR uses static optical path scanning methods based on a fixed ranging module, including micro-electro-mechanical system (MEMS) mirrors, optical phased arrays, and flash illumination. Currently, solid-state LiDAR is still in the pioneering stage and has not yet matured industrially. For samplers of DToF Lidar, there are two options: time-to-digital converter (TDC) and analog-to-digital converter (ADC). Section 3 will discuss the sampler further. The computing unit calculates the ranging results and matches the imaging information to obtain the three-dimensional coordinates of the point cloud.

*2.2. Analysis Method of Ranging Performance of DToF Lidar*

The ranging performance of a lidar, primarily referring to maximum detection range, is directly contingent upon the lidar's optical system and the detection model of the photoelectric signal. In this section, we establish the physical mathematical model of the Lidar's optical system and detection model, and provide a general model for analyzing the maximum detection range of DToF Lidar.

2.2.1. Lidar equation

The light entering the Lidar includes the laser echo reflected from the target surface and background light dominated by solar radiation. Lidar ranging equations which have been widely discussed [23-25] is used to calculate the intensity of laser echo. More general Lidar equations considering measurement direction and target surface orientation will be discussed in our model. As shown in Fig 3(a), the distance between the target and the Lidar is $R$. In most cases (far-field), the laser emission direction is approximately the same as the receiving direction. Angle of laser receiving direction and the horizontal axis is elevation angle $\theta_e$. The one-way transmittance of the laser in the atmosphere $\tau_a = e^{-\alpha R}$, where $\alpha$ is the extinction coefficient including the absorption and scattering of the air. The normal of the target surface is $\vec{N}$, and the angle between the receiving direction and the normal to the target is $\theta$. The unilateral field of view (FoV) of the receiver is $\theta_f = arctan(r_{PD}/f)$, where $r_{PD}$ is the radius of the photosensitive area of the detector (the photosensitive area is circular), $f$ is the focal length of the receiver. The cross-sectional area of the field of view at distance $R$ is $A_r$, and the area of intersection with the target surface is $A_b$. The angle between the direction of solar radiation and the normal to target surface is $\theta_s$. Let the peak power of the laser output by the transmitter be $P_t$. The laser with peak radiant flux $\tau_a P_t$ is incident on the target surface with reflectivity $\rho$. The laser divergence angle is usually in the order of mrad. And the field of view of the Lidar receiver needs to completely cover the laser spot to maximize the signal. Without loss of generality, we assume that the target surface area is larger than the spot area. Therefore, the total radiant flux reflected from the target surface to the hemispheric space is $\rho \tau_a P_t$. According to Lambert's cosine law, the radiant flux on the Lambertian surface is equal to $cos\theta$ times the normal radiant intensity $I_N$. The radiation intensity of the reflected light in the receiving direction is as follows:

$$I_\theta = I_N \cos\theta = \frac{\Phi}{\pi}\cos\theta = \frac{\rho\tau_a P_t}{\pi}\cos\theta \tag{3}$$

The effective aperture (pupil) area of the receiver is $A(\theta_r)$, which is a function of the receiving angle $\theta_r$. The form of the function depends on the scanning method. The solid angle of the aperture to the light spot is:

$$\Omega(\theta_r) = \frac{A(\theta_r)}{R^2} \tag{4}$$

Considering the one way transmittance of the return trip $\tau_a$ and the efficiency $\eta_r$ of the receiving optical system, the echo laser power reaching the photodetector is as follows:

$$P_r = \eta_r \tau_a I_\theta \Omega = \frac{\tau_a^2 \eta_r \rho P_t A(\theta_r)\cos\theta}{\pi R^2} \tag{5}$$

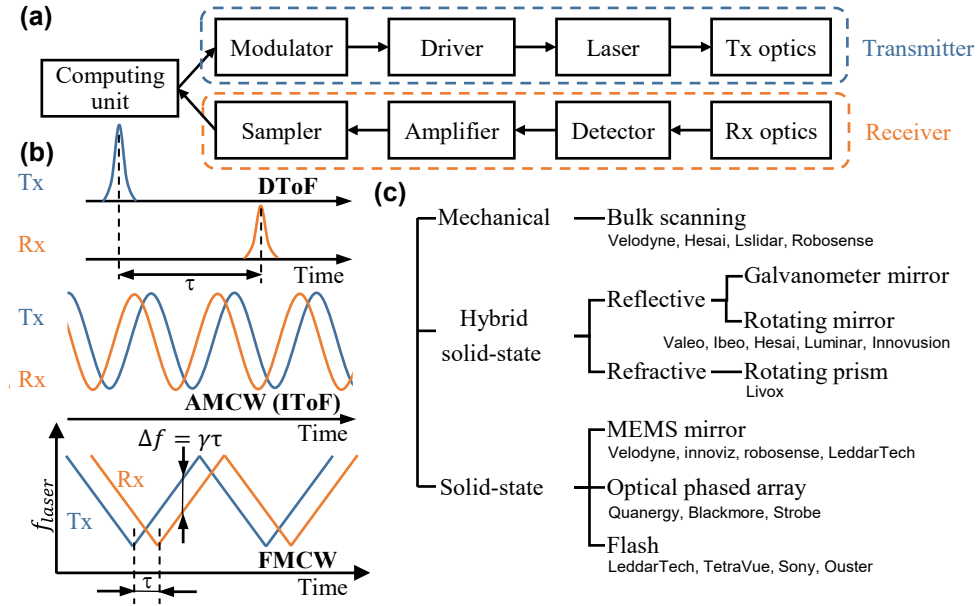

**Figure 2**. Basic knowledge of Lidar (**a**) The components of Lidar (**b**) Three ranging mechanisms for distance measurement: direct time-of-flight (DToF), indirect TOF (amplitude modulated continuous wave, AMCW), frequency modulated continuous wave (FMCW). (**c**) Lidar classification and corresponding companies.

2.2.2. Solar background considerations

The working scene of automotive Lidar is outdoors, so background light usually due to solar radiation that penetrates the earth's atmosphere is superimposed on the laser echo. The thermal radiation of target is small enough to be ignored. The equivalent radiant flux of sun reflected from surfaces within the receiving field of view is:

$$\Phi_s = E_{sun} A_b \cos\theta_s \tag{6}$$

where $E_{sun} = \int_{\lambda_1}^{\lambda_2} E(\lambda) T_{filter}(\lambda) d\lambda$ is the sun equivalent irradiance intensity. $E(\lambda)$ is the solar radiation spectrum (unit: $W/m^2/nm$). $T(\lambda)_{filter}$ is the total transmittance spectrum of the receiver optical system (including window glass, color filter, focusing lens, etc.). $A_b = \frac{A_r}{\cos\theta} = \frac{\pi(Rr_{PD}/f)^2}{\cos\theta}$ is the intersection area of the receiving field of view and the target surface. Similar to Eq. (3), the equivalent radiation intensity of sunlight scattered from the target surface to the receiving direction is as follows:

$$I_{\theta s} = \rho \frac{\Phi_s}{\pi} \cos\theta = E_{sun} \rho R^2 (\frac{r_{PD}}{f})^2 \cos\theta_s \tag{7}$$

Similar to Eq. (5), the optical power generated by solar radiation reaching the photodetector is as follows:

$$P_{rs} = \eta_r \tau_a I_{\theta s} \Omega = E_{sun} \eta_{rs} \tau_a \rho A(\theta_r)(\frac{r_{PD}}{f})^2 \cos\theta_s \tag{8}$$

where $\eta_{rs}$ is the efficiency of the receiving optical system for sunlight.

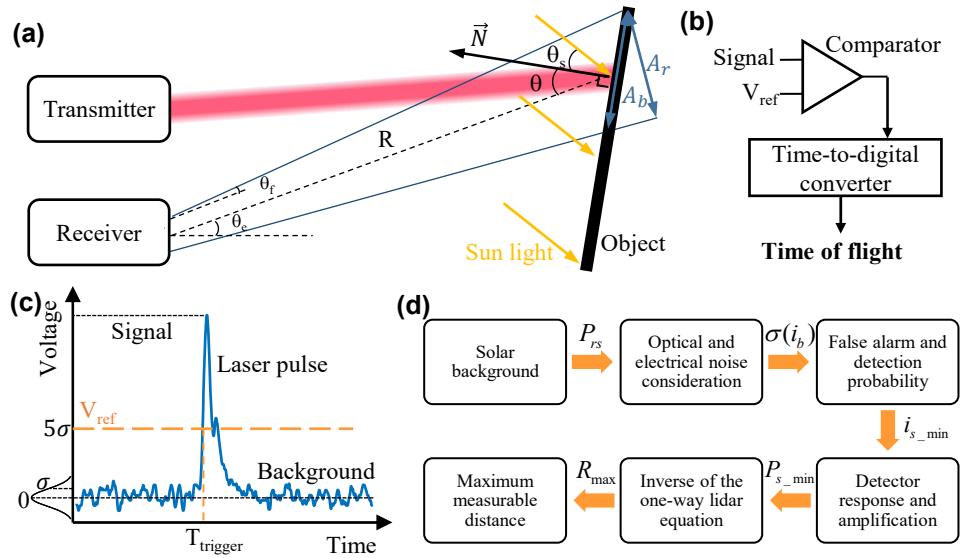

**Figure 3.** Lidar ranging process and calculation of maximum detectable range. (**a**) Schematic diagram of the laser propagation process for a single measurement. The laser is driven to a specific direction by the scanning module and incident on the target surface. Then laser reflected by the target surface and finally received by the receiver. (**b**) Comparator and time-to-digital converter (TDC) record pulse arrival time in photoelectric signal. (**c**) Schematic diagram of TDC triggered by laser pulse. (**d**) The calculation process of the maximum detectable range of DTOF Lidar.

2.2.3. Detection model of photoelectric signal

The optical signal entering the Lidar will be converted into the photoelectric signal through the photodetector and amplifier. As shown in Fig. 3(c), the AC component of photoelectric signal includes the laser echo pulse and the fluctuating signal caused by background light. In the industry, there are two sampling methods, TDC and ADC, utilized for detecting the flight time of laser pulses. The fundamental principle of TDC, as depicted in Fig. 3(b), involves the comparison of the input signal with a reference voltage to ascertain the time difference between two events. ADC is primarily employed to discretize the amplitude values of input photoelectric signal pulses into a series of digital values, representing the signal's amplitude levels. Algorithms can derive more accurate estimations of the laser pulse arrival time and additional information for target identification from the discrete amplitude values provided by ADC. In Lidar systems, ADC sampling rates in the GHz range are typically required, presenting challenges such as large data volume, high chip costs, and high power consumption. Despite TDC's susceptibility to noise and potential timing jitter, it offers advantages such as ease of hardware integration, sufficient accuracy, and scalability in operational principles. It is worth noting that amplitude information of the signal can also be obtained through the extended design of multi-level TDC. Given the current technological level, the data capacity, cost, and power consumption of TDC are more suitable for automotive Lidar. TDC stands as the mainstream solution in the current Lidar industry. Therefore, this paper focuses on analyzing the Lidar based on TDC. Therefore, this paper focuses on the analysis of TDC-based Lidar.

TDC records arrival time of laser echo by voltage threshold triggering. We define the probability that only the background light signal causes the TDC to be triggered as the false alarm probability $P_f$ and the probability that the laser pulse signal causes the TDC to be triggered as the detection probability $P_d$. Without loss of generality, the noise signal of background light obeys a Gaussian distribution $\mathcal{N}(0, \delta(U_b)^2)$. The false alarm probability depends on the SNR of the trigger level $U_{TDC}$ to $\delta(U_b)$, see Eq. (9).

$$P_f = \frac{1}{2} - \frac{1}{2} erf(\frac{U_{TDC}}{\delta(U_b)\sqrt{2}}) \qquad (9)$$

Within the time window $\tau$ of once echo detection, Comparator will make $M = \tau * B_w$ comparisons, $B_w$ is effective bandwidth of the signal amplification circuit. When the laser pulse position is at the end of the time window, the false alarm probability is maximum. At this time, the probability of TDC correctly triggered by the laser pulse is:

$$P_{correct} = \frac{(1-P_f)^{M-1} P_d}{1-(1-P_f)^{M-1}(1-P_d)} \qquad (10)$$

The ratio of the trigger threshold level to the noise $\delta(U_b)$ is referred to as the Threshold Noise Ratio (TNR). In this study, TNR is set to 5 to ensure that the false alarm rate for each comparison is less than $3 \times 10^{-7}$. The limit detection probability is defined as 50%, which means the weakest detectable peak value of the pulse signal $U_s$ is equal to $U_{TDC} = 5\delta(U_b)$. According to Eq. (5), when $\tau = 4\ us$ and $B_w = 100\ MHz$, the correct probability $P_{correct}$ of single ranging is 99.977%. And the maximum detectable range $R_{max}$ corresponds to the weakest detectable peak value $TNR * \delta(U_b)$. In other words, the trigger SNR of Lidar is defined as ratio of the peak value of the signal with laser echo to the standard deviation of signal without laser echo as shown in Eq. (11), and $R_{max}$ is obtained when $SNR = 5$.

$$SNR = \frac{U_s}{\sigma(U_b)} = \frac{i_s(P_r)}{\sigma(i_b(P_{rs}))} \qquad (11)$$

The calculation process of $R_{max}$ can be summarized as Fig 3(d). The algorithm first analyzes the background light intensity entering the Lidar according to Eq. (8), and gets noise photocurrent $\delta(i_b(P_{rs}))$ based on properties of photons, detector and circuit. Then the algorithm calculates the the weakest detectable signal peak $i_{s\_min}$ by letting the trigger SNR be 5. $i_{s\_min}$ can be converted to the weakest detectable laser peak power $P_{s\_min}$ by considering to the detector response and amplifier parameters. Finally, the algorithm Substitutes $P_{s\_min}$ into the Lidar equation (Eq. (5)) to get $R_{max}$.

Therefore, the trigger SNR in Eq. (11) is the key to determine the Lidar ranging performance. The calculation of $P_r$ and $P_{rs}$ in SNR has been discussed in section 3.1 and 3.2, and the analysis of the photocurrent amplitude and noise will be given in section 4 according to the specific photodetector.

*3.1. SNR of DTOF Lidar*

According to Section 3.3, Lidar ranging performance is determined by the trigger SNR which is related to optics, photosensitive devices, and circuits. This section will analyze the DTOF Lidar based on avalanche photodiode and silicon photomultiplier tubes respectively.

There are various options for DTOF Lidar photosensitive devices. Photomultiplier tube (PMT) has single photon detection capability with a very high gain (~$10^6$). However, PMT requires high driving voltage (~kV) and is easily disturbed by magnetic fields and mechanical vibrations. Moreover, the bulk of PMT is also unfavorable for integration into a circuit. In contrast, photodiodes have the advantages of fast response, solid-state and low power consumption. Photodiodes are essentially PN junctions that operates at different reverse voltage. Photodiode (PD) works in the reverse saturation region, avalanche photodiode (APD) works in linear mode (gain~100), single-photon avalanche photodiode (SPAD) and silicon photomultiplier tubes (SiPM) work in Geiger mode (gain~$10^6$) as shown in Fig 4(a). Because PD has no gain capability. When the laser echo is weak, the photocurrent is easily overwhelmed by electrical noise. Photodiodes in Lidar can be APD or SPAD or SiPM. An SiPM is implemented as an array of some single-photon avalanche diodes (SPADs). There are two SiPM forms[40]: analog SiPM (a-SiPM) [26], where the avalanche currents of SPADs are summed in an analog manner; digital SiPM (d-SiPM), where the avalanche currents of SPADs are converted to digital signals and combined using logic trees [41]. Some literatures directly use SPADs [42] to refer to the same meaning as d-SiPM. The array of multiple SiPMs is also called Multi-Pixel Photon Counter (MPPC) or SiPM array or SPADs array. Since the output characteristics of APD and SiPM are different, their SNR need to be discussed separately.

2.3.1. APD based DToF Lidar

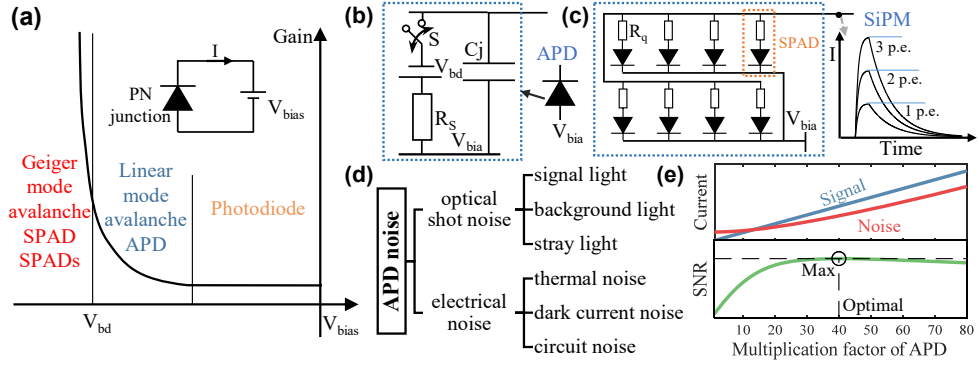

**Figure 4.** Photodetectors for Lidar (**a**) Gain characteristics of PN junction. $V_{bias}$ is bias voltage, $V_{bd}$ is the reverse breakdown voltage. PD works in the reverse saturation region, APD works in linear mode, and SPAD/SiPM work in Geiger mode. (**b**) Equivalent circuit of APD. (**c**) Equivalent circuit of SiPM. (**d**) APD noise overview. (**e**) The relationship between APD trigger signal to noise ratio and multiplication factor.

The APD equivalent circuit is shown in Fig 4(b). We define the signal as the magnitude of the photocurrent generated by the peak power of laser echo as Eq. (14).

$$i_s(P_r) = M_a i_{si} = M_a \frac{e\eta_d P_r}{h\nu} = \frac{e}{h\nu} M_a \eta_d P_r = K_{PD} M_a P_r \qquad (12)$$

where $M_a = (1 - \frac{V_{bias}}{V_{bd}})^{-n}$ is the multiplication factor, $V_{bias}$ and $V_{bd}$ is the APD bias voltage and breakdown voltage, n is a constant related to the detector structure and wavelength of laser. $i_{si}$ is the photocurrent generated before avalanche amplification, $e$ is electron charge constant, $h$ is Planck constant, $\nu$ is laser frequency, $\eta_d$ is quantum efficiency, $K_{PD} = \frac{e}{h\nu}\eta_d$ is defined as the responsivity in A/W before amplification.

The noise is the fluctuation of the photocurrent caused in the optical and electrical processes. We use the root mean square $\sigma$ to quantify noise amplitude. The noise source of APD is shown in Fig 5. The total noise includes signal light shot noise $\sigma_s$, background light shot noise $\sigma_b$, dark current shot noise $\sigma_d$, thermal noise $\sigma_t$, circuit (amplifier) noise $\sigma_c$. The above noises are independent of each other. The influence of stray light can be eliminated through reasonable structure and timing design combined with extinction processing. Therefore, the total noise of APD is as follows:

$$\sigma_{APD} = \sqrt{\sigma_s^2 + \sigma_b^2 + \sigma_d^2 + \sigma_t^2 + \sigma_c^2} \qquad (13)$$

The shot noise currents generated by the echo laser and background light because of the discrete nature of photons and electrons are as follows:

$$\sigma_s^2 = 2e i_{si} B_w M_a^2 F_m = 2e K_{PD} P_r M_a^2 F_m B_w \qquad (14)$$

$$\sigma_b^2 = 2e K_{PD} P_{rs} M_a^2 F_m B_w \qquad (15)$$

where $F_m$ is the excess noise factor of the APD detector, $F_m = k_e M_a + (1 - k_e)(2 - \frac{1}{M_a})$, $k_e$ is the electron ionization rate. $F_m$ can be approximated by $F_m \approx M_a^x$, where $x$ is excess noise index that depends on the semiconductor material, the APD structure and the type of carriers that cause the avalanche effect. $B_w$ is the system effective bandwidth. Due to the discrete nature of electrons, dark current causes fluctuations as follows:

$$\sigma_d^2 = 2e i_{ds} B_w + 2e i_{db} M_a^2 F_m B_w \qquad (16)$$

where the surface dark current $i_{ds}$ does not participate in the multiplication process, and the body dark current $i_{db}$ participates in the multiplication process. Thermal noise (Johnson noise) describes effects of random material-dependent fluctuations on the signal in the receiver.

$$\sigma_t^2 = \frac{4k_B T B_w}{R_l} \tag{17}$$

Where $R_l$ is the load resistance of the detector, $k_B$ is the Boltzmann constant, and $T$ is the effective temperature of the thermal noise source. Defects in devices in actual circuits and coupling interference between devices can also introduce noise $\sigma_c$. Substituting Eq. (15,16,17) into Eq.(15), the noise of Lidar without laser echo can be obtained as:

$$\sigma_{APD}(i(P_{rs})) = \sqrt{2eK_{PD}P_{rs}M_a^2 F_m B_w + 2e(i_{ds} + i_{db}M_a^2 F_m)B_w + \frac{4k_B T B_w}{R_l} + \sigma_c^2} \tag{18}$$

Therefore, the trigger SNR of Lidar with APD is as follows:

$$SNR_{APD} = \frac{i_s}{\sigma_{APD}(i(P_{rs}))} = \frac{K_{PD}M_a P_r}{\sqrt{2eK_{PD}P_{rs}M_a^2 F_m B_w + 2e(i_{ds} + i_{db}M_a^2 F_m)B_w + \frac{4k_B T B_w}{R_l} + \sigma_c^2}} \tag{19}$$

For the denominator of Eq. 19, the first term is optical correlation noise related to responsivity (quantum efficiency), multiplication factor and background light intensity, and the last three terms are electrical noise only related to multiplication factor. Therefore, Eq. 19 can be simplified to

$$SNR_{APD} = \frac{K_{PD}M_a P_r}{\sqrt{a*K_{PD}P_{rs}M_a^{2+x} + b*M_a^{2+x} + c}} \tag{20}$$

Where a b c are constants. Increasing $P_r$ and $K_{PD}$ and reducing $P_{rs}$ can improve the APD trigger SNR. The relationship between multiplication factor and trigger SNR is shown in Fig. 4(e). The appropriate APD multiplication factor can be set to maximize the trigger SNR of APD based Lidar. And it is easy to prove that the optical noise is equal to the electrical noise at this optimal SNR.

2.3.2. SiPM based DToF Lidar

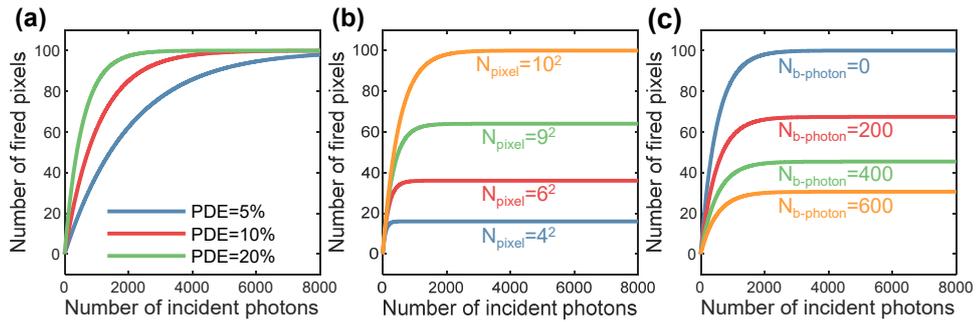

**Figure 5.** SiPM photon response characteristic curve (**a**), SiPM photon response characteristic curve under different photon detection efficiencies ($N_{pixel} = 10^2$ without background light). (**b**), SiPM photon response characteristic curve under different number of pixels ($\eta_{PDE} = 22\%$ without background light). (**c**), SiPM photon response characteristic curve under different background photon numbers ($\eta_{PDE} = 22\%$ $N_{pixel} = 10^2$).

The basic unit of SiPM is SPAD as shown in Fig. 4(c). SPAD has single-photon response capability. Avalanche will not stop once be triggered in SPAD, so the SPAD needs to be used in conjunction with an active or passive quenching circuit (a quenching resistor $R_q$ as shown in Fig. 4(c)). The SPAD is not sensitive during quenching and voltage recovery, this period is called the dead time of SPAD. One SPAD can only be triggered by one photon at a time. To increase dynamic range, SiPM as an array form of SPAD is widely used. When one SPAD is triggered in the SiPM, other SPADs is still able to detect the incident light. The capability of a SiPM to detect photons is quantified by the photon detection efficiency (PDE), which is expressed as:

$$\eta_{PDE}(\lambda, V_{bias}) = \eta_d(\lambda) P_{trigger}(V_{bias}, \lambda) FF(V_{bias}, \lambda) \tag{21}$$

Where $V_{\text{bias}}$ is the bias voltage, $\eta_d(\lambda)$ is the quantum efficiency, $P_{trigger}(V_{\text{bias}}, \lambda)$ is the avalanche trigger probability and $FF(V_{bias}, \lambda)$ is the effective geometric fill factor.

As shown in Fig 4c, the output of SiPM are equivalent to the number of SPADs triggered by judging the interval in which the amplitude is located, so we define the signal as the maximum number of SPADs triggered by the laser echo, and the noise as the random fluctuation of the number of SPADs triggered by background light. The relationship between the number of triggered SPADs and the number of incident photons satisfies the Eq (22) (see Supplemental document for detailed derivation). Since SiPM will not be triggered in the dead time, the number of photons discussed below is the number of photons in one dead time.

$$N_{\text{fired}} = N_{\text{pixel}}(1 - \exp(N_{\text{photon}}(\exp(-\frac{\eta_{PDE}}{N_{\text{pixel}}}) - 1))) \tag{22}$$

Where $N_{\text{fired}}$ is the number of SPADs triggered, $N_{\text{pixel}}$ is the number of SPADs contained in SiPM, $N_{\text{photon}}$ is the total number of incident photons, and $\eta_{PDE}$ is the photon detection efficiency. From Eq. (22), the SiPM photon response characteristics of SiPM depend on $\eta_{PDE}$ and $N_{\text{pixel}}$ as shown in Fig. 5(a,b). The larger the $\eta_{PDE}$, the The better the sensitivity and linearity of response. Increasing of $N_{\text{pixel}}$ will improve the dynamic range of the SiPM, but the power consumption and dark count will increase.

The background light $P_{rs}$ is continuous light, which will continuously trigger $N_b$ SPADs:

$$N_b = N_{\text{pixel}}(1 - \exp(N_{\text{b-photon}}(\exp(-\frac{\eta_{PDE}}{N_{\text{pixel}}}) - 1))) \tag{23}$$

Where $N_{b-photon} = \frac{P_{rs} * \tau}{h\nu}$ is the number of photons of the background light, $\tau$ is the dead time of SPAD. In addition, the thermally excited carriers in the photosensitive region and the carriers tunneled in the high-voltage region will cause avalanche multiplication with certain probability. This is called dark count and is quantified using the dark count rate (DCR) $P_{DCR}$ (unit: cps: count per second). The number of dark counts of SiPM in a dead time is $N_d = N_{\text{pixel}} P_{DCR} \tau$, which is very small ($< 10^{-2}$), so it can be considered that $N_b + N_d$ SPADs are already occupied when the echo pulse is incident. Therefore, the number of SPADs that can be triggered by the signal laser is:

$$N_s = (N_{\text{pixel}} - N_b - N_d)(1 - \exp(N_{\text{s-photon}}(\exp(-\frac{\eta_{PDE}}{N_{\text{pixel}}}) - 1))) \tag{24}$$

Where $N_{s-photon} = \frac{P_r B_{pulse}}{2h\nu}$ is the number of signal photons, $B_{pulse}$ is the full width half maximum of laser pulse in time domain. The response curve of SiPM to incident photons in the presence of background light is shown in Fig 5(c). The stronger the background light, the less dynamic range of SiPM.

According to Appendix A, The optical shot noise of background light causes the fluctuation of SiPM output signal is:

$$\sigma^2(N_b) = N_b \exp(N_{\text{b-photon}}(\exp(-\frac{\eta_{PDE}}{N_{\text{pixel}}}) - 1)) \tag{25}$$

The dark counts follow a Poisson distribution [27], so the dark count noise is as follows:

$$\sigma_d = \sqrt{N_d} = \sqrt{N_{\text{pixel}} P_{DCR} \tau} \tag{26}$$

The effect of the afterpulse and crosstalk on the TDC trigger SNR can be ignored by the amplitude filtering [27]. According to Eq. (24,25,26), the TDC trigger SNR of SiPM is:

$$SNR_{SiPM} = \frac{N_S}{\sqrt{\sigma^2(N_b)+\sigma_d^2}}$$

$$= \frac{(N_{pixel}-N_b-N_{pixel}p_{DCR}\tau)(1-\exp(N_{s\text{-photon}}(\exp(-\frac{\eta_{PDE}}{N_{pixel}})-1)))}{\sqrt{N_b\exp(N_{b\text{-photon}}(\exp(-\frac{\eta_{PDE}}{N_{pixel}})-1))+N_{pixel}P_{DCR}\tau}} \quad (27)$$

In most cases, $N_{pixel}$ is significantly larger than $\eta_{PDE}$, and at the limit range, $N_{pixel}$ is much larger than $N_{s-photon}\eta_{PDE}$. Neglecting small quantities, the above equation can be approximated as:

$$SNR_{SiPM} \approx \frac{N_{s-photon}}{\sqrt{N_{b-photon}}}\sqrt{\eta_{PDE}} = \frac{P_r B_{pulse}}{2\sqrt{hvP_{rs}\tau}}\sqrt{\eta_{PDE}} \quad (28)$$

SiPM trigger SNR is mainly dominated by the optical noise. Increasing $\eta_{PDE}$ and $N_{s-photon}$ and reducing $N_{b-photon}$ can improve the APD trigger SNR. The above discussion is valid when the number of pixels triggering avalanche by background light is far less than the total number of pixels and the laser pulse width is far less than the dead time. However, in DTOF Lidar, a certain level of background light will always be received during the daytime, which does not meet the conditions observed in many single-photon or few-photon SiPM applications. This depends on the specific design. The transient process that triggered SPAD cannot be triggered again during the dead time cannot be ignored. Within the effective bandwidth period, the number of SPADs still in the dead time, excited by the previous period, as well as the number of SPADs excited by background photons and laser photons, along with their fluctuations, will be calculated through simulation. Therefore, when the number of background light photons is relatively large, it is more accurate to simulate the time-domain signal of the SiPM to calculate the SNR of SiPM-based Lidar. When the Lidar designed properly, it is possible to avoid excessive background light entering the system. Moreover, the impact of this phenomenon on systematic results is minimal. Therefore, the analysis results presented above are valid.

## 3. Results and discussion

*3.1. Factors affecting the ranging performance of DToF Lidar*

In accordance with the model depicted in Fig. 3(d) and Eq. (19,27), the formula for calculating the maximum range $R_{max}$ of the APD-based and SiPM-based Lidar is as follows:

$$R_{max\_APD} = (\frac{\cos\theta\tau_a^{1.5}\eta_r}{2\pi\cos\theta_s\sqrt{hv\eta_{rs}\cos\theta_s}})^{1/2}(\frac{\eta_{qe}\rho A(\theta_r)}{E_{sun}Ma^xB_w})^{1/4}(\frac{P_t f}{TNRr_{PD}})^{1/2} \quad (29)$$

$$R_{max\_SiPM} = (\frac{\cos\theta\tau_a^{1.5}\eta_r}{2\pi\cos\theta_s\sqrt{hv\eta_{rs}\tau\cos\theta_s}})^{1/2}(\frac{\eta_{PDE}\rho A(\theta_r)}{E_{sun}})^{1/4}(\frac{P_t B_{pulse} f}{TNRr_{PD}})^{1/2} \quad (30)$$

It is evident that improving the quantum efficiency $\eta_{qe}$ or the photon detection efficiency $\eta_{PDE}$ ($\eta_{qe}$~80%, $\eta_{PDE}$~20% at the current industry level) increasing the receiving aperture $A(\theta_r)$, enhancing the target reflectance $\rho$, reducing the solar irradiance $E_{sun}$, and diminishing the excess noise factor $x$ of APD can enhance the maximum range of DToF Lidar. The specific quantitative relationships can be found in Eq. (28, 29). Therefore, when adjusting design parameters, researchers can calculate the partial derivatives of $R_{max}$ with respect to these design parameters using Eq. (28, 29), to better balance the costs and performance benefits, and select the optimal design solution. Lowering the set value of the threshold-to-noise ratio (TNR) can also increase the range, but at the cost of increased false alarm rate and more noise points in the point cloud. Specific adjustments can be made based on the actual point cloud effect in the scene. Increasing the peak power $P_t$ of the emitted laser and reducing the receiving field of view angle $r_{pd}/f$ can also improve the maximum range, provided that laser safety regulations are met. However, it is important to note that increasing the peak power will reduce the pulse width due to laser safety regulations. This necessitates an increase in the bandwidth $B_w$ of the receiving circuit, which would result in increased noise and costs. Therefore, the design of laser peak power, divergence angle, receiving field of view angle, and circuit bandwidth needs to be comprehensively considered based on the maximum range formula, safety regulations, device characteristics, and costs.

*3.2. Comparison of Ranging Performance between APD and SiPM Lidar*

According to the above model, the ranging performance of APD based DToF Lidar and SiPM based DToF Lidar compared. The DTOF Lidars in this study have identical designs except for the detector. The APD utilized in this paper is the Hamamatsu S12426-02. The parameters of the SiPM are consistent with the macro-pixel of the SPAD array reported by Sony in literature[42], except for the requirement that the photosensitive area matches that of the APD, resulting in a higher number of pixels. The detailed simulation parameters are shown in table 1. Using the physical mathematical models presented in sections 3 and 4, the performance comparison of the signal-to-noise ratio and range based on APD and SiPM DToF lidar is depicted in Fig. 6.

**Table 1.** Parameters used in the calculation of the measurement range

| Property | Variable | Value | Unit |
|---|---|---|---|
| Peak power of laser (905 nm) | $P_t$ | 45 | W |
| Repeat frequency | $f_r$ | 50 | KHz |
| Pulse width | $t_{pulse}$ | 6 | ns |
| Target reflectivity | $\rho$ | 10 | % |
| One way transmission | $\tau_a$ | 98 | % |
| System efficiency | $\eta_r$ | 72.06 | % |
| Effective aperture radius | $r_A$ | 0.025 | m |
| Sunlight illuminance | $lm$ | 100 | Klux |
| Sunlight intensity | $E(\lambda)$ | 29.4 | W/m² |
| Incidence angle of sunlight | $\theta_s$ | 60 | ° |
| Sunlight receiving efficiency | $\eta_{rs}$ | 79.86 | % |
| Focal length | $f$ | 0.03 | m |
| Effective bandwidth | $B_w$ | 167 | MHz |
| APD: S12426-02 | | | |
| Multiplication factor | $M_a$ | 80 | — |
| Radius of photosensitive area | $r_{PD}$ | 0.1 | mm |
| Quantum efficiency | $\eta_{qe}$ | 70 | % |
| Surface dark current | $i_{ds}$ | 0.1 | nA |
| Body dark current | $i_{db}$ | 0.1 | nA |
| Excess noise index | $x$ | 0.3 | — |
| Load resistance | $R_l$ | 10000 | Ω |
| SiPM: sony [42] | | | |
| Number of pixels | $N_{pixel}$ | 20*20 | — |
| Radius of photosensitive area | $r_{PD}$ | 0.1 | mm |
| Photon detection efficiency | $\eta_{PDE}$ | 22 | % |
| Dead time | $\tau$ | 6 | ns |
| Dark count rate | $P_{DCR}$ | 2007 | cps |

In Fig. 6(a), the curve depicts the variation of the trigger signal-to-noise ratio (SNR) with detection distance in the central direction of the scanning field of view (FOV) when the background light is 100 klux (a commonly used industry background light condition). It can be observed from the graph that the trigger signal-to-noise ratio of the Lidar based on APD consistently outperforms that based on SiPM. This is primarily due to the dominance of light noise when the background light is strong, with the trigger signal-to-noise ratio mainly determined by the APD quantum efficiency (70%) and SiPM photon detection probability (22%). It is noteworthy that when the measured distance is relatively close and the echo energy is large, SiPM saturates due to dynamic range limitation, while the dynamic range of APD is determined by specific amplification circuitry. Therefore, additional circuitry design is required to handle high-power echo signals at close distances to prevent their impact on the next measurement cycle. The curve trend indicates that the trigger signal-to-noise ratio is inversely proportional to the square of the distance. Fig. 6b illustrates the trend of the maximum range changing with the scanning direction angle. It can be observed from the graph that the range at the edges of the scanning FOV is lower than that at the center point, mainly due to the different effective receiving apertures in different scanning directions. Fig. 6c demonstrates the comparison of the ranges of the two Lidar under different background light intensities. When the intensity of the background light decreases to a certain value, the range of SiPM exceeds that of APD. The value corresponds to the background light intensity is related to the specific Lidar design and can be calculated specifically using the model provided in this paper. Moreover, the weaker the background light, the more pronounced the advantage of SiPM. Therefore, using SiPM-based solutions has a greater advantage in range in nighttime or indoor scenarios.

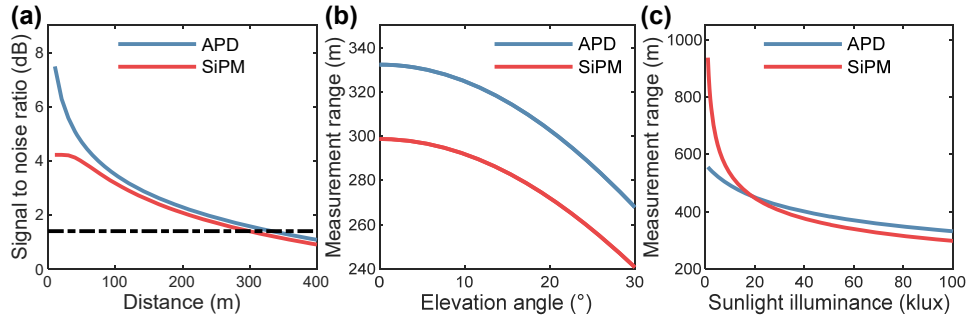

**Figure 6.** Comparison of ranging performance with APD and SiPM as Lidar detectors. (**a**), The relationship between the TDC trigger signal-to-noise ratio and the target distance (elevation angle is 0° and the sunlight illuminance is 100 Klux). The dotted line is the limit TDC trigger signal-to-noise ratio. (**b**), The relationship between the measurement range and the elevation angle when sunlight illuminance is 100 Klux. (**c**), The relationship between the measurement range and the illuminance of sunlight when the elevation angle is 0°.

Controlling the influence of background light is crucial SiPM-based Lidar. From an engineering perspective, background light can be reduced in the following ways: 1) Ensure that the receiving angle of beam is minimized while still capturing the light spot. 2) Minimize the bandwidth of the color filter. 3) Select a laser wavelength that has low energy in the solar radiation spectrum. The conclusions drawn from the range simulation results can be summarized as follows: 1) APD noise can be simply classified as light noise (background light) and electrical noise, while SiPM noise is dominated by light noise. APD can optimize the signal-to-noise ratio by adjusting the multiplication factor to make the light noise equal to the electrical noise. The ultimate noise limit of APD is electrical noise, and SiPM is not affected by electrical noise. It can achieve a higher trigger signal-to-noise ratio after further reducing light noise. Therefore, when working in weak background light scenarios (such as indoor or nighttime scenes), the performance of SiPM-based Lidar is better than that of APD-based Lidar. 2) With the guarantee of dynamic range, the higher the quantum efficiency or photon detection efficiency, the higher the trigger signal-to-noise ratio. 3) Since SiPM has the ability to detect single photons, the design of the optical system needs to reduce the aperture or decrease the receiving angle to ensure that the number of photons in the background light does not significantly affect the dynamic range and response capability of SiPM.

In terms of industry maturity, the APD process is relatively mature, with relatively low cost, mature backend processing circuits, and large-scale production. The SiPM-related industry is becoming increasingly mature, with high photon detection efficiency processes under development, and backend processing circuits and power consumption are being optimized and iterated. Currently, the cost of SiPM is higher than that of APD, but theoretically, SiPM-based Lidar has a longer range and greater potential. From the demand side, the farther the range of the Lidar in automotive applications, the higher the allowable vehicle speed for automatic emergency braking systems, leading to higher efficiency in future traffic systems. Therefore, SiPM-based Lidar is more likely to meet the requirements of automotive Lidar.

## 4. Conclusions

The automotive industry serves as a catalyst for the rapid expansion of the Lidar market, fostering the continual emergence of new participants and innovative products. Lidar is one of the most prominent solutions for achieving a comprehensive perception of the surrounding environment for vehicles. The ranging performance of Lidar is the core of Lidar technology, which involves knowledge of optics, photodetectors, circuits, mechanics and algorithms. This paper elaborates on the calculation methods of the SNR of different Lidars and consolidates the theory of ranging performance of Lidars. Moreover, technological advancements in fields such as lasers, material science, manufacturing and semiconductors will facilitate the further development of Lidar. The theoretical knowledge in this paper can quantitatively analyze how these technological advancements affect the ranging performance of Lidar. This paper presents a reference framework for designing Lidars with various trade-offs between cost and performance, and discusses the potential directions for improving Lidar solutions.

**Author Contributions:** Conceptualization, X.G.; methodology, X.G.; formal analysis, X.G.; investigation, X.G.; writing X.G.; visualization, X.G.; project administration, X.G. All authors have read and agreed to the published version of the manuscript.

**Funding:** This research was funded by SUSTech startup Fund Y01966105

**Data Availability Statement:** Data underlying the results presented in this paper are available from the corresponding author on reasonable request.

**Conflicts of Interest:** The authors declare no conflicts of interest.

## Appendix A

Statistical characteristics of the number of triggered pixels in SiPM will be discussed. In one cycle time corresponding to the effective bandwidth, the total number of incident photons is $q = N_{\text{photon}}$, the total number of SiPM pixels is $N_{\text{pixel}}$. The number k of photons incident to SiPM follows a Poisson distribution:

$$p(k) = \frac{q^k e^{-q}}{k!} \tag{S1}$$

The number of photons incident on each SPDA pixel is $k/N_{pixel}$. The probability that a single pixel is not triggered by $k/N_{pixel}$ photons:

$$p(f=0|k) = \frac{(k\eta_{PDE}/N_{pixel})^f e^{-k\eta_{PDE}/N_{pixel}}}{f!} = e^{-k\eta_{PDE}/N_{pixel}} \tag{S2}$$

The probability that a single pixel is incident with $k/N_{pixel}$ photons and is triggered:

$$p(k, f \neq 0) = p(k) * (1 - p(f=0|k)) = \frac{q^k e^{-q}}{k!}(1 - e^{-k\eta_{PDE}/N_{pixel}}) \tag{S3}$$

Probability of a single pixel being triggered:

$$p(f \neq 0) = \sum_{k=0}^{\infty} \frac{q^k e^{-q}}{k!}(1 - e^{-k\eta_{PDE}/N_{pixel}}) = 1 - \exp(N_{\text{photon}}(\exp(-\frac{\eta_{PDE}}{N_{pixel}}) - 1)) \tag{S4}$$

The expectation of the total number of triggered pixels is:

$$N_{\text{fired}} = N_{\text{pixel}} p(f \neq 0) = N_{\text{pixel}}(1 - \exp(N_{\text{photon}}(\exp(-\frac{\eta_{PDE}}{N_{pixel}}) - 1))) \tag{S5}$$

The standard deviation of the total number of triggered pixels is:

$$\begin{aligned}\sigma^2(N_{\text{fired}}) &= N_{\text{pixel}} p(f \neq 0)(1 - p(f \neq 0)) \\ &= N_{\text{fired}} \exp(N_{\text{photon}}(\exp(-\frac{\eta_{PDE}}{N_{pixel}}) - 1))\end{aligned} \tag{S6}$$

## References

1. Richmond, R.D.; Cain, S.C. *Direct-detection LADAR systems*; SPIE Press: Washington, 2010; Volume 1.
2. Maiman, T.H. Stimulated optical radiation in ruby. **1960**.

3. McManamon, P.F. *Field guide to Lidar*; SPIE Press: Washington, 2015.
4. Flood, M. Laser altimetry: From science to commerical lidar mapping. *Photogrammetric engineering and remote sensing* **2001**, *67*, 1209-1211.
5. Liu, Z.; Zhang, F.; Hong, X. Low-cost retina-like robotic lidars based on incommensurable scanning. *IEEE/ASME Transactions on Mechatronics* **2021**.
6. Areann, M.-C.; Bosch, T.; Lescure, M. Laser ranging: a critical review of usual techniques for distance measurement. *Opt. Eng* **2001**, *40*, 10-19.
7. Winker, D.M.; Hunt, W.H.; McGill, M.J. Initial performance assessment of CALIOP. *Geophysical Research Letters* **2007**, *34*, L19803:19801-19805.
8. Petrovski, A.; Radovanović, M.J.C.M.D. Application of detection reconnaissance technologies use by drones in collaboration with C4IRS for military interested. **2021**, *21*, 117-126.
9. Fahey, T.; Islam, M.; Gardi, A.; Sabatini, R.J.A. Laser beam atmospheric propagation modelling for aerospace LIDAR applications. **2021**, *12*, 918.
10. Vondrak, R.; Keller, J.; Chin, G.; Garvin, J. Lunar Reconnaissance Orbiter (LRO): Observations for lunar exploration and science. *Space science reviews* **2010**, *150*, 7-22.
11. Zhou, G.; Wu, G.; Zhou, X.; Xu, C.; Zhao, D.; Lin, J.; Liu, Z.; Zhang, H.; Wang, Q.; Xu, J.J.I.J.o.A.E.O.; et al. Adaptive model for the water depth bias correction of bathymetric LiDAR point cloud data. **2023**, *118*, 103253.
12. Chen, G.; Hong, L.J.D. Research on Environment Perception System of Quadruped Robots Based on LiDAR and Vision. **2023**, *7*, 329.
13. Zhang, J.; Singh, S. LOAM: Lidar Odometry and Mapping in Real-time. In Proceedings of the Robotics: Science and Systems, Berkeley, CA, 2014; pp. 1-9.
14. Genevois, T.; Horel, J.-B.; Renzaglia, A.; Laugier, C. Augmented reality on lidar data: Going beyond vehicle-in-the-loop for automotive software validation. In Proceedings of the 2022 IEEE Intelligent Vehicles Symposium (IV), 2022; pp. 971-976.
15. Karim, M.R.; Reza, M.N.; Jin, H.; Haque, M.A.; Lee, K.-H.; Sung, J.; Chung, S.-O.J.R.S. Application of LiDAR sensors for crop and working environment recognition in agriculture: A review. **2024**, *16*, 4623.
16. Schwarz, B. Mapping the world in 3D. *Nature Photonics* **2010**, *4*, 429-430.
17. Vinci, G.; Vanzani, F.; Fontana, A.; Campana, S.J.A.P. LiDAR Applications in Archaeology: A Systematic Review. **2024**.
18. Warren, M.E. Automotive LIDAR technology. In Proceedings of the 2019 Symposium on VLSI Circuits, Kyoto, 2019; pp. C254-C255.
19. Kammel, S.; Ziegler, J.; Pitzer, B.; Werling, M.; Gindele, T.; Jagzent, D.; Schröder, J.; Thuy, M.; Goebl, M.; Hundelshausen, F.v. Team AnnieWAY's autonomous system for the 2007 DARPA Urban Challenge. *Journal of Field Robotics* **2008**, *25*, 615-639.
20. Bohren, J.; Foote, T.; Keller, J.; Kushleyev, A.; Lee, D.; Stewart, A.; Vernaza, P.; Derenick, J.; Spletzer, J.; Satterfield, B. Little ben: The ben franklin racing team's entry in the 2007 DARPA urban challenge. *Journal of Field Robotics* **2008**, *25*, 598-614.
21. Leonard, J.; How, J.; Teller, S.; Berger, M.; Campbell, S.; Fiore, G.; Fletcher, L.; Frazzoli, E.; Huang, A.; Karaman, S. A perception‐driven autonomous urban vehicle. *Journal of Field Robotics* **2008**, *25*, 727-774.
22. Vargas, J.; Alsweiss, S.; Toker, O.; Razdan, R.; Santos, J.J.S. An overview of autonomous vehicles sensors and their vulnerability to weather conditions. **2021**, *21*, 5397.
23. Collis, R.T.H. Lidar. *Applied optics* **1970**, *9*, 1782-1788.
24. Kashani, A.G.; Olsen, M.J.; Parrish, C.E.; Wilson, N. A review of LiDAR radiometric processing: From ad hoc intensity correction to rigorous radiometric calibration. *Sensors* **2015**, *15*, 28099-28128.
25. Sandborn, P.A.M. FMCW Lidar: scaling to the chip-level and improving phase-noise-limited performance. University of California, Berkeley, 2017.
26. Villa, F.; Severini, F.; Madonini, F.; Zappa, F. SPADs and SiPMs arrays for long-range high-speed light detection and ranging (LiDAR). *Sensors* **2021**, *21*, 3839.

27. Acerbi, F.; Gundacker, S. Understanding and simulating SiPMs. *Nuclear Instruments and Methods in Physics Research Section A: Accelerators, Spectrometers, Detectors and Associated Equipment* **2019**, *926*, 16-35.
28. Liang, D.; Zhang, C.; Zhang, P.; Liu, S.; Li, H.; Niu, S.; Rao, R.Z.; Zhao, L.; Chen, X.; Li, H.J.n.c. Evolution of laser technology for automotive LiDAR, an industrial viewpoint. **2024**, *15*, 7660.
29. Zhang, C.; Li, H.; Liang, D.J.N.C. Antireflective vertical-cavity surface-emitting laser for LiDAR. **2024**, *15*, 1105.
30. Li, X.; Yang, B.; Xie, X.; Li, D.; Xu, L. Influence of waveform characteristics on LiDAR ranging accuracy and precision. *Sensors* **2018**, *18*, 1156.
31. Li, X.; Zhou, Y.; Hua, B.J.I.T.o.I.; Measurement. Study of a multi-beam LiDAR perception assessment model for real-time autonomous driving. **2021**, *70*, 1-15.
32. Wang, X.; Liu, Y.; Hu, J.; Li, D.; Ma, R.; Zhu, Z.J.I.T.o.C.; Papers, S.I.R. An analog SiPM based receiver with on-chip wideband amplifier module for direct tof lidar applications. **2022**, *70*, 88-100.
33. Wang, X.; Ma, R.; Li, D.; Zheng, H.; Liu, M.; Zhu, Z. A low walk error analog front-end circuit with intensity compensation for direct ToF LiDAR. *IEEE Transactions on Circuits and Systems I: Regular Papers* **2020**, *67*, 4309-4321.
34. Behroozpour, B.; Sandborn, P.A.; Wu, M.C.; Boser, B.E.J.I.C.M. Lidar system architectures and circuits. **2017**, *55*, 135-142.
35. Dai, Z.; Wolf, A.; Ley, P.-P.; Glück, T.; Sundermeier, M.C.; Lachmayer, R.J.S. Requirements for automotive lidar systems. **2022**, *22*, 7532.
36. Holzhüter, H.; Bödewadt, J.; Bayesteh, S.; Aschinger, A.; Blume, H.J.O.E. Technical concepts of automotive LiDAR sensors: a review. **2023**, *62*, 031213-031213.
37. Zhang, X.; Kwon, K.; Henriksson, J.; Luo, J.; Wu, M.C.J.N. A large-scale microelectromechanical-systems-based silicon photonics LiDAR. **2022**, *603*, 253-258.
38. Bamji, C.; Godbaz, J.; Oh, M.; Mehta, S.; Payne, A.; Ortiz, S.; Nagaraja, S.; Perry, T.; Thompson, B.J.I.T.o.E.D. A review of indirect time-of-flight technologies. **2022**, *69*, 2779-2793.
39. Commission, I.E. Safety of laser products-Part 1: Equipment classification and requirements. *IEC 60825-1* **2007**.
40. Gundacker, S.; Heering, A. The silicon photomultiplier: fundamentals and applications of a modern solid-state photon detector. *Physics in Medicine & Biology* **2020**, *65*, 17TR01.
41. Diehl, I.; Hansen, K.; Vanat, T.; Vignola, G.; Feindt, F.; Rastorguev, D.; Spannagel, S.J.J.o.I. Monolithic MHz-frame rate digital SiPM-IC with sub-100 ps precision and 70 μm pixel pitch. **2024**, *19*, P01020.
42. Kumagai, O.; Ohmachi, J.; Matsumura, M.; Yagi, S.; Tayu, K.; Amagawa, K.; Matsukawa, T.; Ozawa, O.; Hirono, D.; Shinozuka, Y. A 189x600 back-illuminated stacked SPAD direct time-of-flight depth sensor for automotive LiDAR systems. In Proceedings of the 2021 IEEE International Solid-State Circuits Conference (ISSCC), Virtual, 2021; pp. 110-112.